\documentclass[usegraphicx,usenatbib,natbib,useAMS]{mn2e}

\usepackage{times}     
\usepackage{aas_macros}
\usepackage{amssymb}   
\usepackage{amsmath}
\usepackage{multirow}
\usepackage{supertabular,multicol}

\def\v90{\mbox{$\Delta v_{90}$}}

\def\vsig{\mbox{$\Delta v_{90}/\sigma_{\mathrm{em}}$}}

\DeclareRobustCommand{\ion}[2]{%
\relax\ifmmode
 \ifx\testbx\f@series
  {\mathbf{#1\,\mathsc{#2}}}\else
  {\mathrm{#1\,\mathsc{#2}}}\fi
 \else\textup{#1\,{\mdseries\textsc{#2}}}%
 \fi}

\title[Exploring DM halos with DLAs]{Exploring galaxy dark matter halos across redshifts with strong
     quasar absorbers}

\author[Christensen, L. et al.]
       {L. Christensen\thanks{lise@dark-cosmology.dk}$^1$, 
         P. M{\o}ller$^2$,
         N.~H.~P.~Rhodin$^1$,
         K.~E.~Heintz$^{3,4}$,
         J.~P.~U.~Fynbo$^4$,\\ 
        $^1$ DARK, Niels Bohr Institute, University of Copenhagen,
         Lyngbyvej 2, 2100 Copenhagen, Denmark\\
        $^2$ European Southern Observatory, Karl-Schwarzschildstrasse
         2, D-85748 Garching bei M\"unchen, Germany\\
         $^3$ Centre for Astrophysics and Cosmology, Science Institute,
    University of Iceland, Dunhagi 5, 107 Reykjav\'ik, Iceland\\
         $^4$ Cosmic Dawn Center, Niels Bohr Institute,
  University of Copenhagen, Lyngbyvej 2, DK-2100 Copenhagen,
  Denmark
}

   \date{Received ; accepted }  

\begin{document}
\maketitle
\label{firstpage}

\begin{abstract}
   Quasar lines of sight intersect intervening galaxy discs or
   circum-galactic environments at random impact parameters and
   potential well depths. Absorption line velocity widths (\v90) are
   known to scale with host galaxy stellar masses, and inversely with
   the projected separation from the quasar line of sight.  Its
   dependence on stellar mass can be eliminated by normalising with
   the emission-line widths of the host galaxies,
   $\sigma_{\mathrm{em}}$, so that absorbers with a range of
   \v90\ values can be compared directly. Using a sample of DLA
   systems at 0.2~$<z<$~3.2 with spectroscopically confirmed host
   galaxies, we find that the velocity ratio \vsig\ decreases with
   projected distances from the hosts. We compare the data with
   expectations of line-of-sight velocity dispersions derived for
   different dark matter halo mass distributions, and find that models
   with steeper radial dark matter profiles provide a better fit to
   the observations, although the scatter remains large. Gas outflows
   from the galaxies may cause an increased scatter, or scale radii of
   dark matter halo models may not be representative for the
   galaxies. We demonstrate by computing virial velocities, that
   metal-rich DLAs that belong to massive galaxy halos
   ($M_{\mathrm{halo}}\approx10^{12}$ M$_{\odot}$) mostly remain
   gravitationally bound to the halos.
\end{abstract}

\begin{keywords}
  quasars: absorption lines --
  galaxies: abundances --
  galaxies: halos --
  galaxies: high-redshift --
  galaxies: kinematics and dynamics --
  cosmology: observations
    \end{keywords}


\section{Introduction}
\label{sect:intro}

A wealth of information about the chemical evolution of the Universe
from low to high redshifts can be obtained from observations of the
strongest hydrogen absorption lines in spectra of luminous sources
such as quasars or gamma-ray bursts. Metallicities have been measured
for hundreds of the strongest intervening absorption systems, the
damped Lyman-$\alpha$ absorbers (DLAs) and sub-DLAs out to redshifts
$z > 5$ \citep[e.g.][]{prochaska03,rafelski14}, revealing a gradual
increase of metallicity with increasing cosmic time. Other notable
important measurements include the evolution of the cosmic neutral
hydrogen density contained in DLAs and sub-DLAs
\citep{noterdaeme12b,zafar13,crighton15,Sanchez-Ramirez16}, and the
evolution of dust properties and dust correction to metallicities
\citep{DeCia18}.

Historically, the nature of DLAs have been debated when only the
absorption-line information was available. Early on, DLAs were
suggested to probe preferentially rotating galaxy discs
\citep{wolfe86} based on the observed velocities and line profiles
with leading edges \citep{prochaska97}. This was supported by models
of disc formation \citep{mo98}, whereas numerical simulations showed
that line profiles could as well be explained by complex gas dynamics
in protogalactic clumps within a hierarchical formation scenario
\citep{haehnelt98}.

DLAs have typically several ($\sim 5-40$) narrow absorption components
identified in their metal absorption lines, with a global velocity
width that is much larger than the intrinsic widths of $\sim
7-20$~km~s$^{-1}$ for each component
\citep[e.g.][]{dessauges03,kulkarni12}. The observed spread of
velocity widths ranging from $\sim$20 and up to $\sim$400 km~s$^{-1}$
suggest a connection to rather massive galaxies, where additional
clouds in the galaxy halos besides components that arise in the galaxy
discs contribute to the global and diverse absorption line profiles
\citep{wolfe05}. Particularly, the presence of outflows from galaxies
can explain higher velocity components seen in some DLAs
\citep{bouche13}. Numerical simulations have shown that average DLAs
(dominated by low-metallicity systems) arise in halos of
preferentially faint, low-mass galaxies
\citep{nagamine07,cen12,rahmati14,bird15}, but the large cosmological
simulations do not have a sufficient resolution to distinguish
individual components that contribute to a single DLA system.

Understanding the nature of DLAs hinges on the knowledge of the
connection between DLAs in absorption and their host galaxies detected
in emission. At redshifts $<1$ the success rate of identifying galaxy
counterparts is relatively high since the galaxies are brighter
\citep[e.g.][]{chen05,peroux11a}. More recent observations with
modern, sensitive instruments have provided a breakthrough by
spectroscopically confirming in emission the host galaxies associated
with higher-redshifts DLAs
\citep[e.g.][]{fynbo10,peroux11a,noterdaeme12,krogager13,fynbo13,neeleman18}.
To date, the majority of detected galaxies are preferentially
associated with metal-rich DLAs. The numerous failures of searches for
DLA galaxy counterparts and upper detection limits can be explained
with a simple model that involved a luminosity-metallicity relation,
where non-detections mostly belong to low-metallicity DLAs, and hence
low-mass and low-luminosity galaxies \citep{krogager17}.

Besides the directly measurable metallicities, DLAs themselves contain
additional information about the parent galaxy. The velocity widths of
metal absorption lines spanning 90\% of the integrated optical depths
\citep[defining \v90 in][]{prochaska97,wolfe98} scale with absorber
metallicity in a relation that evolves with redshift
\citep{ledoux06,prochaska08,neeleman13}. Moreover, this relation can
be interpreted as a relation between DLA metallicity and host galaxy
halo mass \citep{moller13}, reflecting the known luminosity-selected
galaxy mass-metallicity relation \citep{tremonti04,maiolino08}.  By
measuring the stellar masses of the host galaxies, we now know that
DLAs indeed follow a mass-metallicity relation
\citep{christensen14,augustin18,rhodin18}.

One of the remaining pieces of the puzzle is how the DLA velocity
widths and metallicities observed at some random impact parameters,
defined as the projected distance between the host galaxy and quasar
line of sight, spanning from a few- and up to $\sim$100 kpc, are
affected by the host itself. DLAs have been suggested to arise either
in infalling pristine gas from the intergalactic medium or in outflows
from the host galaxies \citep{bouche13,peroux16}, or simply neutral
halo gas extending far from the galaxies.

Located at projected distances between a few and up to $\sim$100 kpc
from the host galaxies, DLAs must experience the gravity from the
matter distribution of the host galaxy, both from baryonic matter at
scales defined by the stellar components, and the dark matter (DM)
potential.  Using both quasars and gamma-ray bursts (GRBs) to trace
DLA systems with average spatial offsets between GRB locations and the
host galaxy centres, \citet{arabsalmani15} suggest that the location
relative to the galaxy affects both velocities and metallicities with
a radial dependence that cancel each other. In \citet{moller19} we
investigate the dependence of the projected spatial location on the
velocity and metallicity information from the DLA itself by comparing
quasar and GRB-DLAs, and find a scaling relation between DLA
velocities, host galaxy masses and DLA impact parameters. We find
evidence for an increasing velocity width (measured by \v90) in close
projection to the host galaxy, but also a dependence that scales with
the velocity dispersion ($\sigma_{\mathrm{em}}$) of the host galaxy
itself.

In this paper, we analyse the velocities in DLAs and their host
galaxies as probes of the gravitational potential at the location of
the DLA. Different models of DM distributions in galaxy halos give
rise to variations of the projected line-of-sight velocity dispersions
as a function of radius from the central galaxy
(Sections~\ref{sect:sigma_los} and \ref{sect:scale_models}) depending
on the host galaxy halo masses. We compare the velocities with
numerical simulations in Section~\ref{sect:halo_vir}, and show
remarkable correspondence between simulations and the data. Section
\ref{sect:summary} presents the summary.

\section{DLA velocity widths and galaxy scaling relations}
\label{sect:DLA_velo}

Models of galaxy formation can be used to predict the rotational
velocities of galaxies \citep[e.g.][]{mo98}.  Unfortunately,
observations of DLAs do not provide information of the circular
velocity of the DM halos or detailed transversely resolved velocity
information. DLAs only probe velocity components of individual clouds
along a single line of sight at a random impact parameter from the
centre of the galaxy, and rotation curves of the host galaxies have
been measured only in a few instances
\citep[e.g.][]{chen05,peroux11a,bouche13}.  Whereas the gaseous
structure that comprise a DLA system might be large, individual clouds
that contribute to the DLAs can be much smaller, with sizes as small
as 0.1 pc \citep{krogager16}. However, much larger structures covering
coherent sizes of 100 kpc have also been reported for DLAs
\citep{ellison07}. Since DLAs contain several components separated in
velocity space, where each one is located inside the potential well of
the galaxy, we may use the full DLA system as a probe of the host halo
velocity dispersion.

Logically, more massive galaxies in more massive halos give rise to
higher velocity dispersion at a fixed distance, so in order to compare
galaxies spanning a large range in masses, we also need to know the
masses of the galaxies and halos themselves.  Because \v90 scales with
the metallicity of the galaxy \citep{ledoux06,neeleman13} it follows
from the mass-metallicity relations that a scaling between \v90 and
the stellar mass of the parent galaxy exists too \citep{moller13}.  In
addition, the galaxies themselves obey scaling relations with
velocities. For example, the relation between the stellar mass and the
velocity width ($\sigma_{\mathrm{em}}$) of strong emission lines, the
so-called stellar-mass Tully-Fisher relation \citep{kassin07} is well
known. For various galaxy samples this relation is found to be
redshift invariant at $z\lesssim3$ \citep{christensen17}.

The existence of these scaling relations allow us to compare DLA
systems with a large range of \v90 arising from galaxies with a large
range in stellar masses.  We compile data from the literature for a
sample of observed DLA systems, where host galaxies have been detected
in emission and emission-line velocities are reported \citep[see
  details in][]{moller19}. Table~\ref{tab:halomass} presents the
sample of 21 DLAs at redshifts $0.2< z<3.2$ for which we have
information of both \v90 and $\sigma_{\mathrm{em}}$. We also include
an additional 11 identified DLA galaxies with measured stellar masses
in Table~\ref{tab:halomass}, which will be used to investigate how the
velocities correlate with halo masses.

\begin{figure}
\begin{center}
  \includegraphics[width=8.6cm]{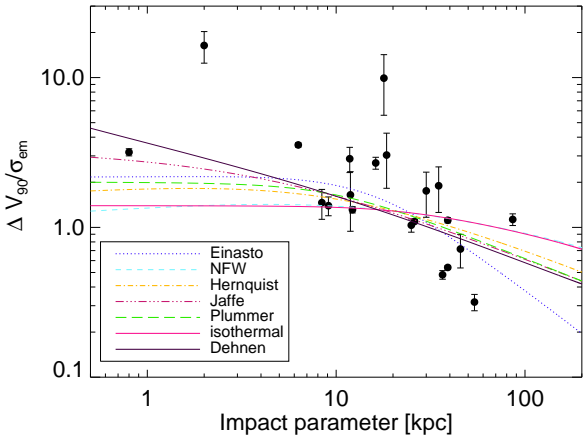}
\end{center}
\caption{Measured velocities for DLAs and their host galaxies as a
  function of impact parameters. The lines represent
  $\sigma_{\mathrm{los}}$ from various DM mass distribution
  profiles. Based on computed real values from the models we argue
  that \v90$\approx\sigma_{\mathrm{los}}$.  All models have a scale
  radius of 10 kpc, and are  normalised to provide the smallest
    $\chi^2$ with respect to the data.}
\label{fig:plot1}
\end{figure}

In Fig.~\ref{fig:plot1} we illustrate the measured ratio
$\v90/\sigma_{\mathrm{em}}$ as a function of impact parameters for the
DLAs.  In \citet{moller19} we argue that this relation arises as
metallicity traces the local gravitational potential at the location
of the DLA.

Following on that result we here investigate the underlying reason for
this scaling relation.  The gravitational potential that is
  dominated by dark matter at large distances can be computed from
  theoretical models of its mass distribution, and here we aim to test
  a wide range of such models against this new set of observational
  constraints.

\begin{table*}
  \centering
\begin{tabular}{llllrrrrrr}
\hline
\hline
Quasar   & $z_{\mathrm{abs}}$ &  $z_{\mathrm{em}}$ & $b$ &  \v90 & $\sigma_{\mathrm{em}}$  & log$M_*$ & log$M_{\mathrm{halo}}$  & $r_s$($z=z_{\mathrm{abs}}$) &
References \\
          &  &  & [kpc] & [km~s$^{-1}$] & [km~s$^{-1}$] & [log
  M$_{\odot}$] & [log M$_{\odot}$] &  [kpc] \\
\hline
0439--433   & 0.1012 & 0.1010 & 7.2  &   275  & ---        & 10.01$\pm$0.02 & 11.56 &  17.8 & 23\\ 
 0151+045   & 0.1602 & 0.1595 & 18.5 &   152  &  50$\pm$20 &  9.73$\pm$0.04 & 11.43 &  15.9 & 2,3 \\
0738+313    & 0.2212 & 0.2222 & 20.3 &    60  & ---        &  9.33$\pm$0.05 & 11.26 &  13.6 & 9,24,29 \\ 
 1127--145  & 0.3127 & 0.3121 & 17.5 &   123  & ---        &  8.29$\pm$0.08 & 10.80 &  8.8  & 9,24 \\ 
 0827+243   & 0.5247 & 0.5263 & 38.4 &   188  & ---        & 10.09$\pm$0.15 & 11.76 &  21.8 & 9,24,29  \\
 2328+0022  & 0.65179& 0.65194& 11.9 &    92  &  56$\pm$24 & 10.62$\pm$0.35 & 12.21 &  33.2 & 4,1 \\
 2335+1501  & 0.67972& 0.67989& 27.0 &   104  & ---        &  9.83$\pm$0.21 & 11.64 &  19.4 & 4,6,1 \\ 
 1122--1649 & 0.6819& 0.68249 & 25.6 &   144  & ---       &  9.45$\pm$0.15 & 11.33 &  14.5 & 4,24 \\ 
 1323--0021 & 0.71612& 0.7171 & 9.1  &   141  & 101$\pm$14 & 10.80$\pm$0.10 & 12.47 &  42.3 & 5 \\ 
 1436--0051A& 0.7377 & 0.73749& 45.5 &    71  &  99$\pm$25 & 10.41$\pm$0.09 & 12.03 &  27.9 & 4,6 \\
 0153+0009  & 0.77219& 0.77085& 36.6 &    58  & 121$\pm$8  & 10.03$\pm$0.13 & 11.78 &  22.0 & 4,1 \\
 0152--2001 & 0.7798 & 0.78025& 54.0 &    33  & 104$\pm$13 &    ---         &   --- &  ---  & 7\\
 1009--0026 & 0.8866& 0.8864  & 39.0 &    94  & 174$\pm$5  & 11.06$\pm$0.03 & 13.18 &  58.4 & 6,8,9\\ 
 1436--0051B& 0.9281& 0.92886 & 34.9 &    62  &  33$\pm$11 & 10.20$\pm$0.11 & 11.91 &  24.7 & 4,6\\ 
 0021+0043  & 0.94181 &0.9417 & 86.0 &   139  & 123$\pm$11 &  ---           & ---   &  ---  & 10,1\\ 
 0452--1640 & 1.00630 & 1.0072& 16.0 &    70  & ---        &  9.1$\pm$0.2   & 11.31 &  14.0 & 27,28,1  \\
 0302--223  & 1.00945&1.00946 & 25.0 &    61  &  59$\pm$6  &  9.65$\pm$0.08 & 11.60 &  18.4 & 8,9 \\ 
 2352--0028 & 1.03917&1.0318  & 12.2 &   164  & 125$\pm$6  &  9.4$\pm$0.3   & 11.47 &  16.2 & 6,27,28,1\\ 
 2239--2949 & 1.82516& 1.8260 & 20.8 &    64  &  ---       &  9.81$\pm$0.50 & 11.77 &  20.4 & 25,16 \\  
 2206--1958 & 1.91999& 1.9220 & 8.4 &    136  &  93$\pm$21 &  9.45$\pm$0.30 & 11.57 &  16.8 & 9,11\\
 1228--1139 & 2.19289& 2.1912 & 30.0 &   163  &  93$\pm$31 &    ---         & ---   & ---   & 12,13\\
 1135--0010 & 2.2066 & 2.2073 &  0.8 &   168  &  53$\pm$3  &    ---         & ---   &  ---  & 14\\ 
 0124+0044  & 2.26223& 2.2620 & 10.9 &   142  &  ---       & 10.16$\pm$0.14 & 12.01 &  23.3 & 16,1\\ 
 2243--603  & 2.3298 & 2.3283 & 26.0 &   173  &  158$\pm$5 & 10.10$\pm$0.10 & 11.98 &  22.2 & 15,16,26\\
 2222--0946 & 2.3542 & 2.3537 &  6.3 &   174  &   49$\pm$2 &  9.62$\pm$0.12 & 11.69 &  16.8 & 17,9,22 \\
 0918+1636  & 2.4121 & 2.4128 &  2.0 &   344  &   21$\pm$5 &   ---          &   --- & ---   & 18 \\
 1439+1117  & 2.41802& 2.4189 & 39.0 &   338  &  303$\pm$12& 10.74$\pm$0.18 & 12.50 &  35.3 & 19,20\\
 0918+1636  & 2.5832 & 2.58277& 16.2 &   288  &  107$\pm$10& 10.33$\pm$0.08 & 12.15 &  24.3 & 18,9 \\
 0139--0824 & 2.6771 & 2.6772 & 13.0 &   168  &  ---       &  8.23$\pm$0.20 & 10.90 &   7.3 & 16,1 \\ 
 0528--250  & 2.8110 & 2.8136 &  9.1 &   304  &  ---       &  8.79$\pm$0.15 & 11.22 &   9.5 & 9,24\\ 
 2358+0149  & 2.97919& 2.9784 & 11.8 &   135  &   47$\pm$9 &   ---          &   --- &   --- & 21\\ 
 2233+1318  & 3.14930& 3.15137& 17.9 &   228  &   23$\pm$10&  9.85$\pm$0.14 & 11.85 &  15.9 & 11,9\\
\hline
\end{tabular}
\caption{Properties of the DLAs and their host galaxies. The first
  column lists quasar names representing those from original
  discoveries, and therefore frequently referring to coordinates from
  B1950 equinox. Other columns present: absorber redshift (2), galaxy
  emission redshift (3), impact parameter for the identified DLA host
  (4), absorber velocity dispersion (5), galaxy emission-line velocity
  dispersion (6), stellar mass for the identified host (7). Halo
  masses in Column 8 are computed from halo abundance matching methods
  \citep{moster13}, and scale radii (column 9) $r_s$ using
  equation~\ref{eq:rs2}. References for stellar masses, and DLA
  velocity widths, redshifts and impact parameters: [1] VLT/UVES or
  X-shooter archive, this work, [2] \citet{christensen05}, [3]
  \citet{som15} [4] \citet{rhodin18}, [5] \citet{moller18}, [6]
  \citet{meiring09}, [7] \citet{rahmani18}, [8] \citet{peroux11b}, [9]
  \citet{christensen14}, [10] \citet{peroux16}, [11]
  \citet{weatherley05}, [12] \citet{ellison05}, [13]
  \citet{neeleman18}, [14] \citet{noterdaeme12}, [15]
  \citet{bouche13}, [16] Rhodin et al. in prep., [17] \citep{fynbo10},
        [18] \citep{fynbo13}, [19] \citet{rudie17}, [20]
        \citet{srianand08}, [21] \citet{srianand16}, [22]
        \citet{krogager13}, [23] \citet{neeleman16}, [24]
        \citet{kanekar14}, [25] \citet{zafar17}, [26]
        \citet{ledoux06}, [27] \citet{peroux13}, [28]
        \citet{augustin18}, [29] \citet{chen05}}
\label{tab:halomass}
\end{table*}

\section{Line-of-sight velocity dispersions in halos}
\label{sect:sigma_los}

To compare the observations with predictions of velocity dispersions
of galaxy halos, we need to compute the expected radial profiles. Far
from the stellar disc of the galaxy and the baryonic components, the
mass contribution is dominated by dark matter. Assuming a density
distribution of the DM one can compute the line-of-sight velocity
dispersion, $\sigma_{\mathrm{los}}(R)$, as a function of projected
radial distance, $R$, from the galaxy. This quantity should be
equivalent to measuring a velocity dispersion as a function of the
impact parameter as provided by the observations of \v90. For an
isotropic distribution
\citep{binney87,merritt87,hernquist90,dehnen93}, this can be computed
as

\begin{equation}
  \sigma_{\mathrm{los}}^2(R) = \frac{2}{\Sigma(R)} \int_{R}^{\infty}
  \sigma_r^2(r) \rho(r) \frac{r}{{\sqrt{r^2-R^2}}} dr, 
\label{eq:sigma2} 
\end{equation}
where the normalisation depends on the projected surface density given by
\begin{equation}
  \Sigma(R) = 2 \int_{R}^{\infty} \rho(r) \frac{r}{\sqrt{r^2-R^2}} dr.
\label{eq:sigma} 
\end{equation}
$\sigma_r(r)$ is the radial velocity dispersion, which can be derived
from the Jeans equation as
\begin{equation}
  \sigma_r^2(r) = \frac{1}{\rho(R)} \int_{R}^{\infty}  \rho(r')
  \frac{GM(r')}{r'^2} dr',
\label{eq:sigma_r} 
\end{equation}
with the mass profile as function of radius being
\begin{equation}
  M(r) =  \int_{0}^{r} 4\pi r'^2 \rho(r') dr'.
\label{eq:mass} 
\end{equation}

All equations above depend on the radial density profile, which can be
parameterised as a double power-law
\begin{equation}
\rho(r) =  \frac{\rho_0}{(r/r_s)^{\gamma}[1+(r/r_s)^{\alpha}]^{(\beta-\gamma)/\alpha}} ,
\label{eq:density} 
\end{equation}
where $r_s$ is a scale radius and $\rho_0$ the central density.
Various combinations of [$\alpha,\beta$, $\gamma$] give the known profiles from
\citet{hernquist90} [1,4,1], \citet{jaffe83} [1,4,2],
\citet{plummer11} [2,5,0], and NFW \citet{navarro97} [1,3,1] and an
isothermal profile [2,3,0] \citep{binney87}.  Other special cases of
DM profiles are suggested by \citet{dehnen93}:
\begin{equation}
\rho(r) =  \frac{(3-\gamma)M}{4\pi}\frac{r_s}{r^{\gamma}(r+r_s)^{4-\gamma}}
\label{eq:density3} 
\end{equation}
where the factor $\gamma$ can be chosen to give a steeper inner
profile than other density profiles for $\gamma~>~2.5$. In addition to
this suite of halo models, we also investigate the Einasto profile
\citep{einasto65} parameterised as
\begin{equation}
  \rho(r) = \exp(-2n((r/r_s)^{1/n}-1)).
\label{eq:density2} 
\end{equation}

All these radial DM profiles include the most common and
  classical halo models used in theoretical works. As the density
  profiles have varying slopes the differences in the projected
  line-of-sight velocity dispersions are apparent at either very small
  or large radial distances. 

Line-of-sight radial velocity dispersion profiles,
$\sigma_{\mathrm{los}}(R)$, for the various density distributions are
illustrated in Fig.~\ref{fig:plot1}. To compare models with the data
points, we initially use a common scale radius of $r_s=10$ kpc for all
models, $n=1$ defining a relatively steep slope for the Einasto
profile, and $\gamma=2.75$ for the \citet{dehnen93} model to represent
a very steep inner density profile. The velocity dispersion profiles
are normalised to provide the smallest $\chi^2$ values to fit the
data.

To evaluate whether $\sigma_{\mathrm{los}}$ for a halo gives a value
representative of \v90 for a DLA, we compute the absolute value of
$\sigma_{\mathrm{los}}$ by setting the scale radius $r_s=17$ kpc, and
from abundance matching in Sect.~\ref{sect:scale_models}, we compute
the median DLA halo mass from Table~\ref{tab:halomass} to be
$M_{\mathrm{halo}}=10^{11.7}$ M$_{\odot}$. 
This gives $\sigma_{\mathrm{los}}\sim 100-130$ km~s$^{-1}$ depending
on the chosen DM halo mass profile at $r=r_s$, while the median
measured \v90 = 141 km~s$^{-1}$ in Table~\ref{tab:halomass}. Therefore
it is a good approximation to set $\v90 \approx
\sigma_{\mathrm{los}}$.  The median stellar mass of the DLA galaxies
in Table~\ref{tab:halomass} is $M_*=10^{9.8}$~M$_{\odot}$, for which
the stellar-mass Tully-Fisher relation in \citet{christensen17}
implies an emission-line velocity dispersion of
$\sigma_{\mathrm{em}}\approx 60$ km~s$^{-1}$. With these simple
considerations, we expect the curves in Fig.~\ref{fig:plot1} to lie
around $\vsig\approx 1.7-2.2$ at $b\sim17$ kpc in good agreement with
the measured data points and their spread.

\section{Galaxy scale radii in halo models}
\label{sect:scale_models}

\subsection{Dark matter halo scale radii}
Since this analysis involves galaxies with very different stellar
masses over a range of redshifts, a direct comparison as done in
Section \ref{sect:sigma_los} is too simplified since the scale radii
of the galaxies are not the same. In order to place the galaxies in
the same system and compare to models, it is relevant to scale the
observed impact parameters with the galaxies' scale radii. These scale
radii depend on the galaxy mass and dark matter concentration
parameters, which again depend on redshifts.

The total stellar masses are known for some of the galaxies in
Table~\ref{tab:halomass}, and are computed by fitting spectral energy
distributions (SEDs) to template spectra created with a range of star
formation histories, stellar ages, reddenings, and metallicities. In
all SED fits, a Chabrier initial mass function has been adopted. We
refer the reader to \citet{christensen14} and \citet{rhodin18}, where
SED fits to DLA galaxies are explained in detail. To compute the DM
halo masses, we use the formalism from halo abundance matching methods
described in \citet{moster13}. Halo abundance matching techniques
generally find that the fraction of stellar to dark matter mass peaks
around $M_{\mathrm{halo}}=10^{12}$ M$_{\odot}$ with a weak redshift
evolution. For each of the DLA galaxies with known stellar masses we
compute halo masses listed in Table~\ref{tab:halomass}, taking into
account the redshift evolution from abundance matching models. The
halo masses correlate with the DLA metallicity, following the scaling
between the host galaxy stellar-mass and DLA metallicity. Halo masses
of $\sim10^{12}$ M$_{\odot}$ for these metal-rich absorbers agree with
the high bias factors from cross-correlations with the Lyman-$\alpha$
forest for metal-rich systems and consequently large halo masses
\citep{perez-rafols18}.

The next step is to derive the scale radii. The halo mass within the
virial radius of the galaxy ($r_h$) can be described as
\begin{equation}
M_{\mathrm{halo}} = \frac{4}{3}\pi r_h^3 \Delta_c(z) \rho_c(z),
\label{eq:halo_rad}
\end{equation}
where $\Delta_c(z)$ is the overdensity and $\rho_c(z)$ is the critical
energy density in a flat universe at redshift, $z$
\begin{equation}
\rho_c(z) = \frac{3 H(z)^2}{8 \pi  G}.
\label{eq:rhoc}
\end{equation}
The Hubble parameter evolves as \(H(z)=H_0 E(z) \) with $E(z)^2 =
\Omega_{0,m}(1+z)^3+\Omega_{0,\Lambda}$. In this paper, we use a flat
cosmology with $\Omega_{0,\Lambda}=0.727$, $\Omega_m=0.273$, and
$H_0=70.4$ km~s$^{-1}$~Mpc$^{-1}$ \citep{komatsu11}. The overdensity can
be parametrised as
\begin{equation}
\Delta_c(z) = 18 \pi^2 +82[\Omega(z)-1] -39[\Omega(z) -1]^2
\label{eq:delta}
\end{equation}
\citep{bryan98,posti14},
where 
\begin{equation}
\Omega(z) = \Omega_{0,m}(1+z)^3/E(z)^2.
\end{equation}

The scale radius and halo radius are connected through the
concentration parameter, $c~=~r_h/r_s$. Numerical simulations
\citep{mo04} have shown that the concentration parameter depends on
the halo mass
\begin{equation}
c(M) = 11 \Big(\frac{M_{\mathrm{halo}}}{10^{12}h^{-1} \mathrm{M_{\odot}} }\Big)^{0.15}.
\label{eq:cm}
\end{equation}
More recent numerical simulations find a different dependence
\citep{klypin11}
\begin{equation}
c(M) = 9.6 \Big(\frac{M_{\mathrm{halo}}}{10^{12}h^{-1} \mathrm{M_{\odot}} }\Big)^{-0.075}.
\label{eq:cm2}
\end{equation}

Combining equations~\ref{eq:halo_rad}, \ref{eq:rhoc}, \ref{eq:delta}
and \ref{eq:cm}, we get
\begin{equation}
r_s^3 = M_{\mathrm{halo}}^{0.55}\frac{3}{4\pi 11^3}\frac{1}{\rho_c(z)
  \Delta_c(z)} \big(10^{12} h^{-1} \mathrm{M_{\odot}}\big)^{0.45}
\label{eq:rs}
\end{equation}
or alternatively with equations~\ref{eq:halo_rad}, \ref{eq:rhoc},
\ref{eq:delta} and \ref{eq:cm2}, we get
\begin{equation}
r_s^3 = M_{\mathrm{halo}}^{1.225}\frac{3}{4\pi 9.6^3}\frac{1}{\rho_c(z)
  \Delta_c(z)} \big(10^{12} h^{-1} \mathrm{M_{\odot}}\big)^{-0.225}.
\label{eq:rs2}
\end{equation}
Whether we use equation~\ref{eq:cm} or \ref{eq:cm2}, for the redshift 
dependence of the concentration parameters, the differences in the 
output results are insignificant relative to the scatter of the data points.

\begin{figure}
\begin{center}
\includegraphics[width=8.6cm]{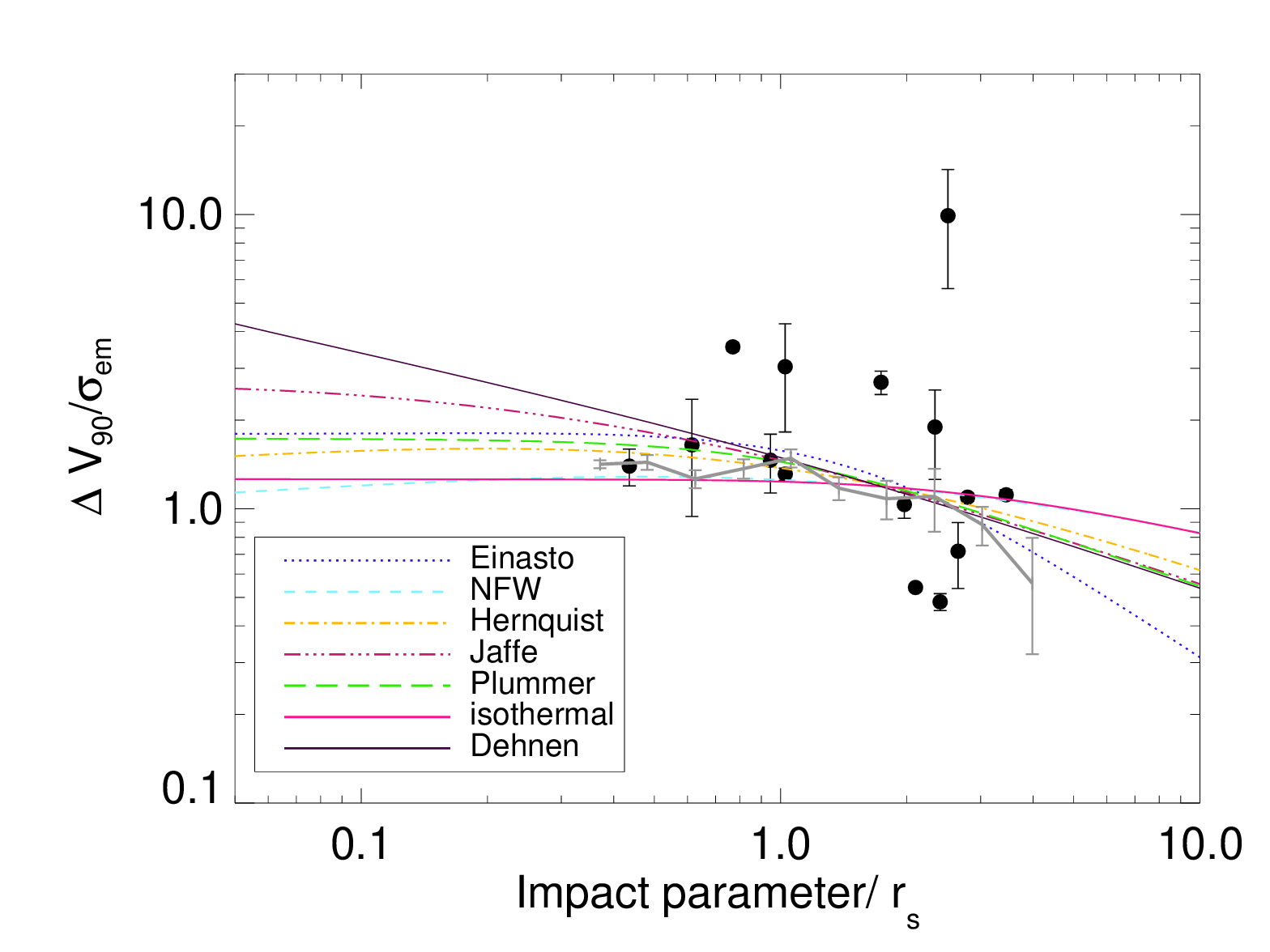}
\includegraphics[width=8.6cm]{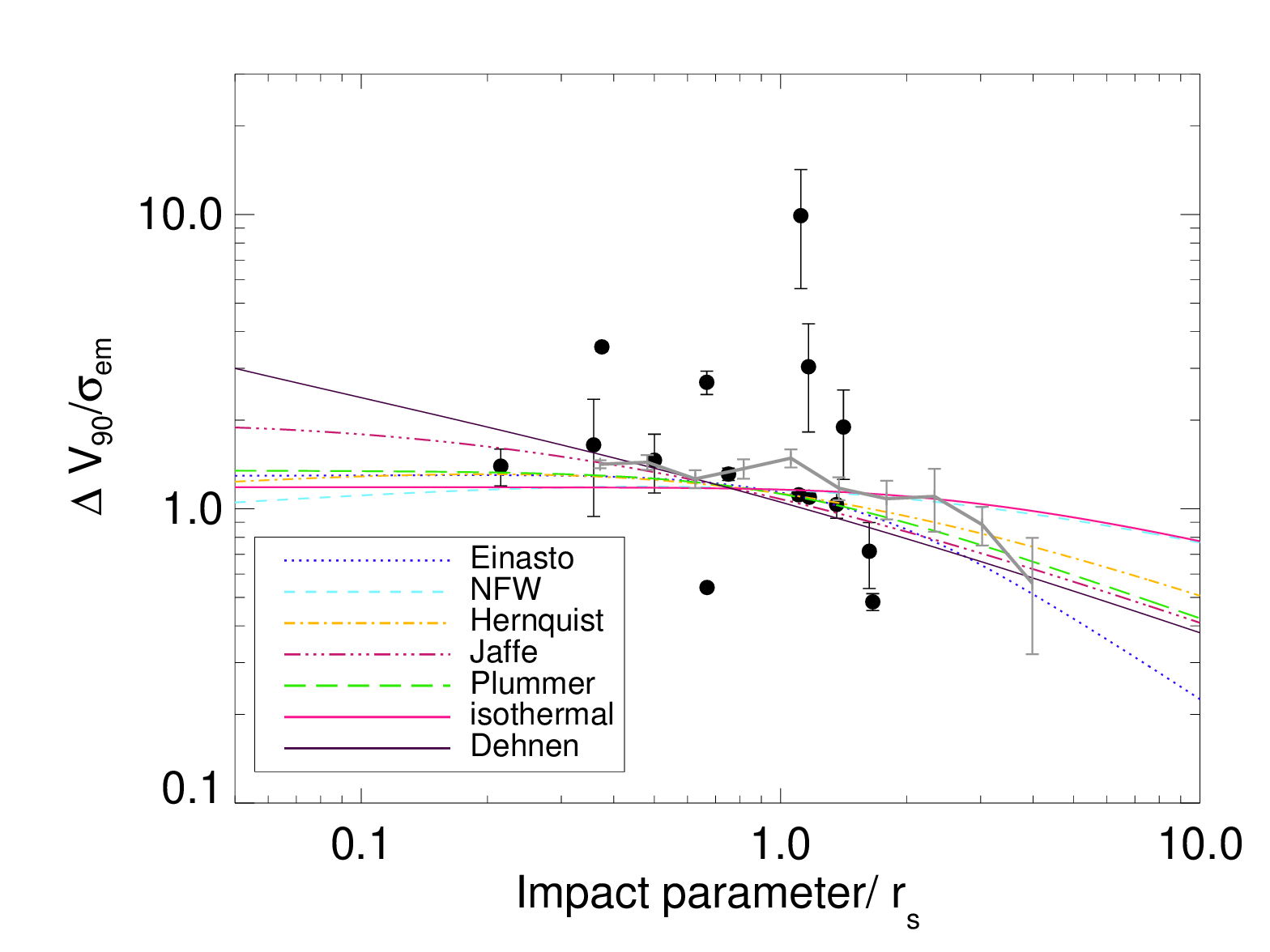}
\end{center}
\caption{Plot of the data versus line of sight velocity dispersions
  for different DM profiles. All data point are scaled in the x-axis
  with their scale radii. The upper panel ignores the change of the
  scale radii with redshifts and assumes $z=0$ using
  equation~\ref{eq:rs2}. All models have $r_s=1$ kpc and have been
  normalised to provide the minimum chi square residuals with respect
  to the DLA data points.  The grey line presents the radial velocity
  dispersion distribution of stars in the Milky Way halo
  \citep{battaglia05}, which has been arbitrarily normalised to
  $\sigma_{\mathrm{em}} = 90$ km~s$^{-1}$.  {\it The lower panel}
  includes a redshift dependent scale radius (equation~\ref{eq:rs_z}),
  and the halo models have been normalised to provide the minimum
  $\chi^2$ value with respect to the DLA data points.  }
\label{fig:plot2}
\end{figure}

The concentration parameter decrease with redshift at $z<2$ roughly as
\begin{equation}
      c(M)(z)= \frac{c(M)_{z=0}}{(1+z)^{0.75}}
\end{equation}
 \citep{klypin11}, while at higher redshifts the decrease in
the concentration parameter levels off \citep[see also][]{zhao09}.
At higher redshifts ($z>3$) and for very large halo masses
($M_{\mathrm{halo}}>10^{13}$ M$_{\odot}$) the trend changes and the
concentration parameter starts to increase. However, apart from a
single object (DLA1009--0026), the halos and redshifts involved in
this analysis do not reach this regime, so we take a very simplistic
approach.  To reflect the inversion of $c(M)$ for the massive host of
DLA1009--0026 we assume that $c\sim$7, similar to the also relatively
massive host of DLA1323--0021 that has a slightly lower redshift.
Including a redshift dependence, with \(r_s\propto (1+z)^{0.75}\),
equation~\ref{eq:rs2} becomes
\begin{equation}
r_s^3 = M_{\mathrm{halo}}^{1.225}\frac{3}{4\pi 9.6^3}\frac{1}{\rho_c(z)
  \Delta_c(z)} \big(10^{12} h^{-1} \mathrm{M_{\odot}}\big)^{-0.225} (1+z)^{2.25}.
\label{eq:rs_z}
\end{equation}

The scale radii for the DLA galaxies are listed in column 9 in
Table~\ref{tab:halomass}. Fig.~\ref{fig:plot2} illustrates the result,
where impact parameters are normalised by their scale radii,
reflecting the same radial dependence across all measured DLAs.  The
DM halo profiles have been normalised to produce the minimum $\chi^2$
value for the measured data points.  Table~\ref{tab:chi} lists the
  values of $\chi^2/dof$ for each model fit. Models with a flatter DM
  density profile slope, such as the Isothermal or NFW profiles have
  worse fit compared to the steeper profiles for Dehnen, Jaffe or
  Einasto models.

\begin{table}
  \centering
\begin{tabular}{lll}
\hline
\hline
Model  & $\chi^2/dof$(Fig.~2)  & $\chi^2/dof$(Fig.~1)\\
\hline 
Einasto    & 149  &  58 \\
NFW        & 161  & 119 \\
Hernquist  & 148  &  83 \\
Jaffe      & 134  &  62 \\
Plummer    & 147  &  71 \\
Isothermal & 161  & 120 \\
Dehnen     & 127  &  57 \\
\hline
\end{tabular}
\caption{Reduced $\chi^2$ values for the different model fits. Second
  column lists $\chi^2/dof$ for the model fits to Fig.~\ref{fig:plot1}
  where a uniform scale radius of $r_s=10$ kpc has been assumed, and
  where the impact parameters are not normalised to the computed scale
  radii for individual galaxies.
}
\label{tab:chi}
\end{table}

There may be a complication in combining numerous diverse halos with
only a single line of sight probed in each case. To compare with the
velocity dispersion distribution observed in the Milky Way, we
over-plot halo star velocity dispersion measurements
\citep{battaglia05}.  In order for the Milky Way to be placed in the
same scaled system, we computed the scale radius $r_s=31$ kpc at $z=0$
based on the stellar mass $\log M_*$=10.8 M$_{\odot}$ for the Milky
Way \citep{licquia14}, and assume $\sigma_{\mathrm{em,MW}}=90$
km~s$^{-1}$. As seen, the MW stars follow well the DM potential given
by either the Einasto, Plummer or Hernquist models, whereas the DLA
data points exhibit a larger scatter.

\subsection{Adding baryonic mass components}

The halo models addressed above only contain dark matter. However,
the baryonic components in the form of stellar mass and gaseous
material also contribute to the potential and therefore also to the
projected line-of-sight velocity dispersion.

To evaluate the contribution to $\sigma_{\mathrm{los}}$ we add a
baryonic component to a halo with a mass $M_{\mathrm{halo}}=10^{11.7}$
M$_{\odot}$, derived from the median halo mass in our sample.  We add
an exponential disc galaxy profile with a density profile $\rho(r)
\propto \exp(-z/r_z) \exp(-r/r_d)$, where the disc scale length $r_d$
is varied between 0.1 and 1 kpc. A very small disc scale length is
used because the halo models use scale radii of 1~kpc.  To compute the
mass distribution, the disc height, $r_z$, is assumed to be always
equal to 100 pc. Such a disc model is clearly very simplified, but
serves the purpose of describing the changes of
$\sigma_{\mathrm{los}}$ when including baryons.  The total baryonic
mass is varied from $10^{10}-10^{11}$ M$_{\odot}$, representing a
baryonic fraction of $2-20$\%. The radial mass profile of the galaxy
is added to equation~\ref{eq:mass}. Since the radial velocity
dispersion $\sigma(r)$ depends on the dynamics of the baryonic disc,
which is not known, we assume it to be equal to the rotational speed
of the baryons alone. Finally, this is added in quadrature to
$\sigma_{\mathrm{los}}$ for the DM halo to compute the combined
$\sigma_{\mathrm{los,DM+baryons}}$.

\begin{figure}
\begin{center}
\includegraphics[width=8.6cm]{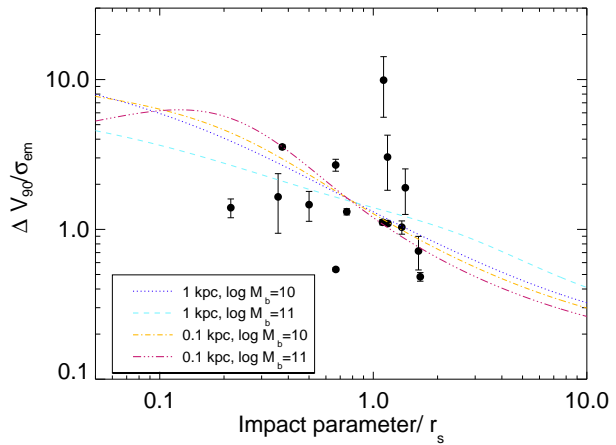}
\end{center}
\caption{Illustration of the change of $\sigma_{\mathrm{los}}$ as a
  result of adding baryonic components to the DM halo profiles. For
  the DM mass distribution, the Jaffe profile is used as an example, and
  various galaxy models with exponential scale radii of 0.1 and 1 kpc
  and baryonic masses of $\log M_b=10-11$ M$_{\odot}$ are added as
  explained in the legend.}
\label{fig:plot3}
\end{figure}

The changes for the computed $\sigma_{\mathrm{los}}$ profiles are
illustrated in Fig.~\ref{fig:plot3}. When adding a massive disc
component, there is a large contribution from baryons to
$\sigma_{\mathrm{los}}$ around the disc scale length, while for less
massive discs, the difference from the pure DM halo model is less
pronounced.

The stellar masses of the DLA galaxies in our sample span almost three
orders of magnitude from log $M_*=8.3-11.1$, but we do not know the
contribution from cold neutral gas to the entire galaxy mass.  As DLA
galaxies are by definition absorption selected, and therefore
sensitive to the amount of neutral gas present, the baryonic mass may
also preferentially be in the form of gas that has not yet been
processed in star formation, or does not form stars at the time of
observations. We therefore test the expected $\sigma_{\mathrm{los}}$
in the case of a gas mass 10 times that of the stellar mass.
Such high values of atomic gas to stellar mass are found in local
low-mass galaxies selected from the 21~cm {\sc Alfalfa} survey
\citep{huang12}. Similarly, recent 21~cm emission studies of local
absorption selected (low stellar mass) galaxies have revealed H{\sc i}
mass to stellar mass ratios of $5-100$ \citep{kanekar18b}, and in
$z\sim0.6$ absorption selected galaxies, high molecular gas fractions
have been detected \citep{moller18,kanekar18}. Whether higher redshift
DLA systems that probe more massive galaxies also have such high H{\sc
  i} gas mass fractions, is not supported by any current observations
but remains to be verified by future observations.
A realistic \ion{H}{i} disc is larger in size relative to the stellar
disc. Adding such a massive, extended \ion{H}{i} disk causes the
velocity dispersion profile to be flatter with a bump around the
chosen scale radius. This is not consistent with the measured data
points, and moreover with a fraction of $M_{\mathrm{HI}}/M_*\sim100$,
the total baryonic mass is similar to or even higher than the total DM
halo mass, in conflict with the commonly accepted DM to baryonic mass
fraction.

In conclusion, the models that include a stellar and gaseous
components in addition to the DM halo mass are able to explain the
radial dependence of \vsig. For reasonable baryon to DM fractions, the
contributions from baryons do not play significant role compared to
the dominant DM contribution at large impact parameters.

\section{Halo virial velocities}
\label{sect:halo_vir}
\subsection{Comparison with numerical simulations}

Having derived halo virial masses, we compare the measured \v90 values
with the distribution of halo virial velocities and \v90 values
derived from numerical models of DLAs \citep{haehnelt98,bird15}.  In
the simulations, random lines of sight through galaxy halos are drawn
and simulated spectra of DLAs are created with radiative transfer
models. Then metal absorption line widths are measured for galaxies
with known virial masses and virial velocities \citep{bird15}. The
ratio between velocities $\v90/V_{\mathrm{vir}}$ peaks around 0.9,
with a significant spread in the ratio.

In order to compare the observed DLA systems with simulations, we need
to determine the halo virial velocity for each of the 26 DLA galaxies
in Table~\ref{tab:halomass} for which we know the stellar and halo
masses. We assume the simple relation that the rotational velocity at
the radius $r_{\mathrm{vir}}=r_{200}$, where the overdensity
$\Delta_c$ is 200 times that of the cosmic value, is equal to the
circular velocity \(V^2_c = G M_{\mathrm{halo}}
r_{\mathrm{vir}}^{-1}\). The halo mass at this radius is determined by
combining equations \ref{eq:halo_rad} and \ref{eq:rhoc}:
\(M_{\mathrm{halo}} = 100 r_{\mathrm{vir}}^3 H^2(z) G^{-1}\), and the
halo virial velocity can be computed as
\begin{equation}
  V^3_{\mathrm{vir}}= 10 G M_{\mathrm{halo}} H(z).
\end{equation}

\begin{figure}
\begin{center}
  \includegraphics[width=8.6cm]{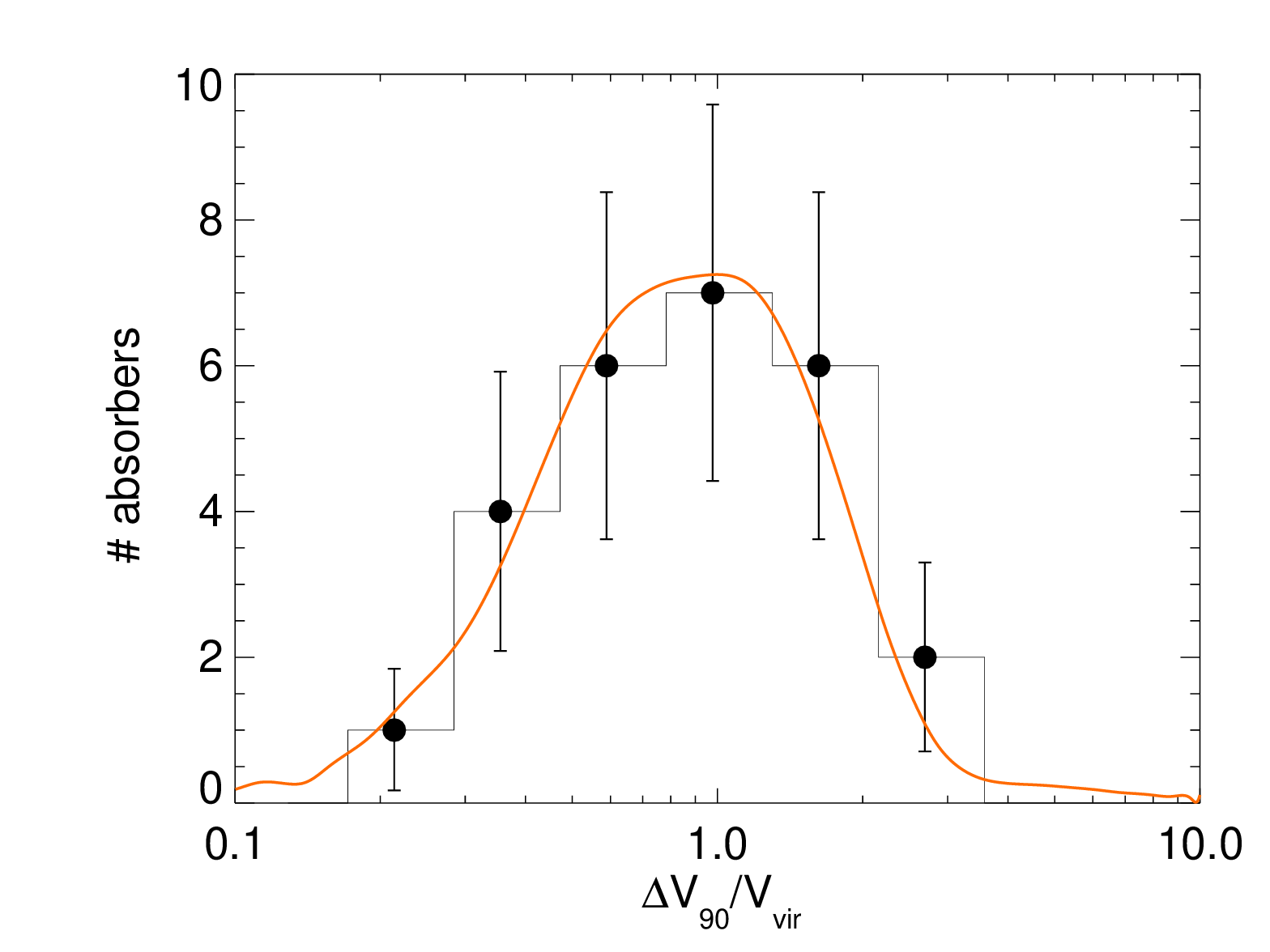}
  \includegraphics[width=8.6cm]{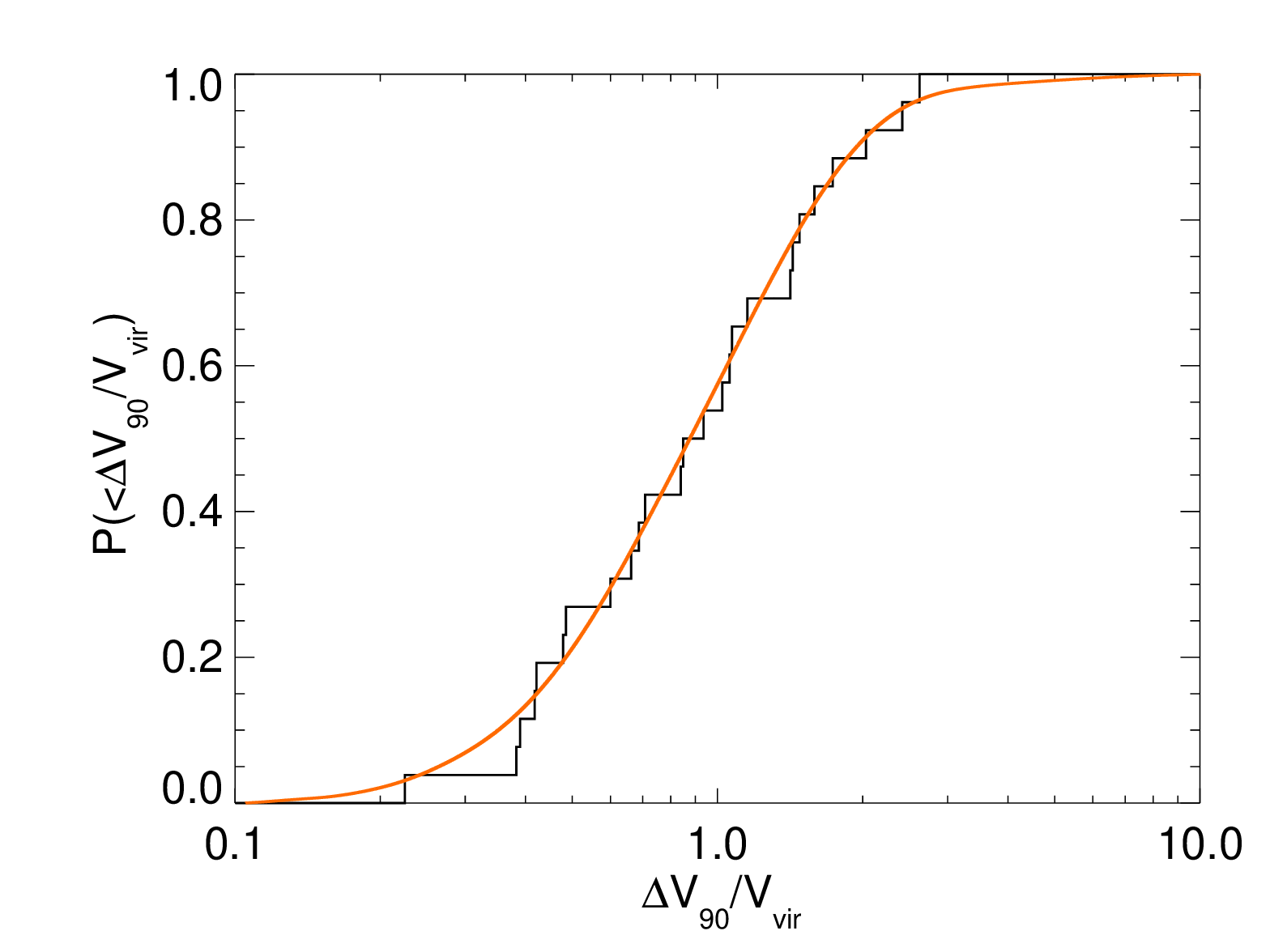}
\end{center}
\caption{The top panel shows a histogram of the fractional velocities
  for 26 DLAs with known stellar- and halo masses, versus simulated
  DLAs at $z=3$ in \citet{bird15} illustrated by the orange curve. The
  models are not fit to the data, but simply scaled, and the vertical
  errorbars denote 68\% confidence levels given by Poisson statistics
  in the small number regime \citep{gehrels86}.  The bottom panel
  shows the cumulative distribution of the same data, and a KS-test
  gives a probability of $P=0.98$ that the two distributions are the
  same.}
\label{fig:plot4}
\end{figure}

Comparing our observed data to the models, Fig.~\ref{fig:plot4} shows
a remarkable similarity. The histogram of $\v90/V_{\mathrm{vir}}$ has
errorbars representing 68\% confidence regions for small-number
Poisson statistics \citep{gehrels86}. Testing the full cumulative
distribution with a two-sided Kolmogorov-Smirnov (KS) test gives a
probability of $P=0.98$ that the two distributions are drawn from the
same underlying sample.

We note that the simulated DLA data are measured at a single redshift
of $z=3$, whereas the data points belong to a compilation of DLAs at
all redshifts between 0.1 and 3.2. We checked for a possible redshift
dependence by splitting the observed data in two redshift intervals at
$z<1.0$ and $z>1.0$, for which the KS test probabilities are $P=0.64$
and $P=0.38$, respectively. For the high redshift sample, the observed
galaxies show higher velocities than the simulated ones.
Still, the cumulative distribution shapes remain similar to that of
the full sample, so that the evidence for a systematic shift with
redshift is weak.

\citet{bird15} find that the average halo virial velocity for DLAs is
70 km~s$^{-1}$, while the halos we have analysed here have higher
velocities ranging from $60-420$ km~s$^{-1}$ and a median of
$V_{\mathrm{vir}}=145$ km~s$^{-1}$, simply because the DLA hosts
detected to date are dominated by more massive and luminous galaxies
belonging to relatively metal-rich DLAs.  In addition, impact
  parameters are not presented for the simulations, but if a radial
  dependence of \v90\ also exist in simulations, it would imply that
  lower \v90\ values were more common because the cross section at
  larger radii is higher. The excellent agreement illustrated in
  Fig.~\ref{fig:plot4} could therefore be a coincidence. For a proper
  comparison with numerical simulations we need to know the stellar-
  or halo masses of hosts for each of the individual simulated DLAs,
  as well as the impact parameters used for determining \v90.

\begin{figure}
\begin{center}
  \includegraphics[width=8.6cm]{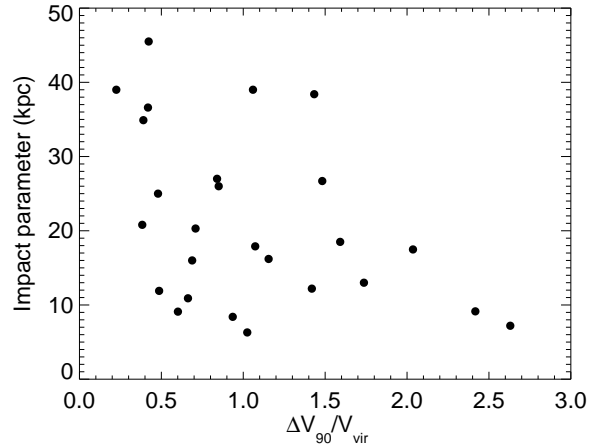}
\end{center}
\caption{The impact parameter as a function of the velocity ratios
  $\v90/V_{\mathrm{vir}}$ suggests a dependence that the
  higher fractions are only found at small radial separations from the
  host galaxies.}
\label{fig:plot5}
\end{figure}

Even though we do not know the spatial position of the individual
components along the line of sight that give rise to the full
absorption line profile in observations, the correspondence does give
credit to the interpretation that DLA absorption line widths trace the
host halo potential.  Fig.~\ref{fig:plot5} illustrates the dependence
of the impact parameter on the derived $\v90/V_{\mathrm{vir}}$
fraction. Although there is no clear scaling relation, no DLA systems
with a high \v90 velocity width relative to $V_{\mathrm{vir}}$ is
found at large impact parameters, indicating that the DLA systems are
gravitationally bound to the halos.  To check if redshifts and the
local gravitational potential play an important role for
$\v90/V_{\mathrm{vir}}$ in simulations, it is necessary to compare the
observed data with simulated DLAs at a range of redshifts with known
impact parameters from simulated galaxies that better match the
observed masses of the DLA galaxies.  Such comparisons between models
and simulations will be the aim of a future study.

\subsection{DLA systems are bound to the host halos}
Having computed the halo mass distributions, we can proceed to ask if
the DLA systems are bound to the halos, or whether their velocities
are sufficient to allow them to escape from the parent galaxy
potential \citep[see also][]{moller19}. By computing the relative
velocity offsets from the DLA absorption redshift and galaxy emission
redshifts listed in Table~\ref{tab:halomass}, we can compare with the
escape velocities at the DLA impact parameters.

\begin{figure}
\begin{center}
  \includegraphics[width=8.6cm]{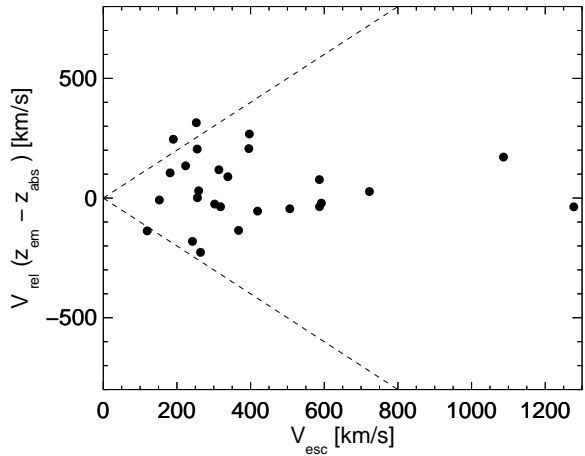}
\end{center}
\caption{Relative velocity offset between absorption and emission
  redshifts compared to the computed escape velocities from a
  Hernquist mass-distribution model at the position equal to the
  impact parameter. The data points that lie above and below the
  straight dashed lines correspond to systems that have sufficient
  velocities to escape the potential well of the host galaxy.}
\label{fig:plot6}
\end{figure}

First we scale the halo mass profile to the total mass of the DLA DM
halo plus the galaxy stellar mass. Then we find the enclosed mass at
the DLA impact parameter and compute the escape velocity at that
position. Fig.~\ref{fig:plot6} illustrates that from the 26 DLA-galaxy
pairs only three DLA systems (DLA0738+313, DLA1127--145, and
DLA0827+243) have relative velocities that allow them to escape the
host galaxy potential. However, these three relative velocities are
just barely above the escape velocities by 55, 18 and 62 km~s$^{-1}$,
respectively. The illustration in Fig.~\ref{fig:plot6} uses a
Hernquist density distribution, but another density profile will not
change the conclusion. For the remaining 23 DLAs, and even if the DLAs
arise in galaxy outflows, the gas does not have sufficiently large
velocities to allow it to escape the potential well of the galaxies.

We can only measure the radial velocity component along the line of
sight from the absorption and emission redshifts, meaning that any
tangential velocity differences are ignored. However, this correction
is expected to be small compared to the computed escape
velocities. The correction is of the order of a few $10 - 100$
km~s$^{-1}$ computed from Equation~\ref{eq:sigma_r} to represent the
tangential velocity (dispersion) on the plane of the sky.

\section{Discussion and summary}
\label{sect:summary}
In this paper we use absorption and emission lines in DLA systems and
their host galaxies to probe the gravitational potential of the host
galaxy halo at the random line of sight to the background quasars
through the intervening galaxy. We compare observed DLAs with
predictions of velocity dispersions from a range of dark matter
models.  The velocity offsets between DLA systems and host galaxies
compared to escape velocities at the measured impact parameters
demonstrate that DLA clouds mostly remain gravitationally bound to the
host galaxies.  We furthermore compare the ratio of \v90 to virial
  velocities with numerical simulations, and while the match is good,
  it could be a coincidence as the observations with detected DLA
  hosts systematically include more massive galaxies than the
  simulations.

\subsection{DLAs as probes of galaxy halos}

The absorption line width, \v90, has previously been proven to scale
with the DLA metallicity and also with the stellar mass of the host
galaxy. As a consequence, metal-rich DLA systems also belong to
massive halos. In a sample of 26 metal-rich DLA systems with
metallicities in the range $-1.2<\mathrm{[M/H]}<0.6$ with
spectroscopically confirmed host galaxies, we compute their
corresponding halo masses to lie in the range
$10^{10.8}<M_{\mathrm{halo}}<10^{13.2}$~M$_{\odot}$ with a median of
$10^{11.7}$~M$_{\odot}$. Rotation curves of three DLA galaxies that
are also included in our sample yielded similarly high halo masses
\citep{peroux13}. These high values of halo masses are in excellent
agreement with the high bias factor measured by cross-correlating DLAs
with the Ly$\alpha$ forest, which implies that those metal-strong DLAs
arise in massive halos $M_{\mathrm{halo}}=10^{12}$~M$_{\odot}$, while
metal-weaker systems arise in halos that are two orders of magnitude
less massive \citep{perez-rafols18}.

The \v90 values lie between $\sim 20-400$~km~s$^{-1}$ corresponding to
a large dynamical range of galaxy masses. In order to compare
velocities in galaxies with masses spanning almost 3 orders of
magnitude, we therefore normalise this velocity width by the velocity
dispersion of the host galaxy traced by its emission lines,
$\sigma_{\mathrm{em}}$. We investigate the radial dependence of this
dimensionless parameter \vsig\ for DLAs and their host galaxies, and
compare with line-of-sight velocity dispersions expected from various
DM model distributions. The DLA data suggest a steep radial dependence
of the distribution out to a distance of $\sim$60 kpc from the host.
 While the different halo mass distributions give quite different
  predictions for the line-of-sight velocity dispersions at either
  very large or small impact parameters, the currently known data
  sample does not allow us to rule out any of the models.

Extra mass-components from baryons, which dominate at low impact
parameters, give rise to steeper or flatter profiles depending on the
(baryonic) scale radii.  However, in order to explain the steepness of
the relations and the scatter of data points, we have to modify the
baryonic component for each DLA system individually such that no
global scaling relation is obvious. In comparison, the velocity
dispersions of halo stars in the Milky Way apparently can be described
to follow pure DM models without the need to add a baryonic component.

\subsection{Spread of \vsig\ measurements}

Several effects could contribute to the observed scatter of the data
points relative to that predicted by the various DM halo models.
Outflows from galaxies would increase both \v90 and
$\sigma_{\mathrm{em}}$ as the latter also scales with galaxy
star-formation rates \citep{kruhler15,christensen17}. Outflows will
therefore cause minor increase in the scatter of the
\vsig\ ratio. Additionally, some DLAs are known to arise in galaxy
groups \citep{kacprzak10,peroux17,fynbo18}, and dynamical interactions
between group members and their overlapping halos are likely to affect
the kinematics of the absorbers, as detected in the complex
inter-group gas kinematics associated with the $z=0.313$ DLA towards
Q1127--145 \citep{chen19}.

The DM halo models presented in this work assume spherical symmetry,
which does not need to be the case. In galaxy clusters for example,
velocity anisotropies along different directions have been measured
\citep[e.g.][]{wojtak09}. How this can affect individual halos such as
those probed by DLAs is not clear, but any anisotropy cannot explain
the very large values of \vsig\ in Fig.~\ref{fig:plot2}.

When comparing the \vsig\ distributions in Fig.~\ref{fig:plot1}
  that were not scaled according to DM halo models \citep[see
    also][]{moller19} to the scaled models in Fig.~\ref{fig:plot2}, it
  is not evident that the more advanced theoretical models provide
  better fits to the data. Table~\ref{tab:chi} shows that the unscaled
  data have less scatter compared to the predicted velocities from all
  DM distributions. When computing the scale radii, $r_s$, we rely on
  abundance matching of halo to stellar mass ratios. However, studies
  of weak gravitational lensing, Tully-Fisher relations, and stellar
  kinematics have demonstrated that individual galaxy halo masses can
  be under-predicted by abundance matching by as much as a factor of
  10 \citep{leauthaud12}. Scale radii would accordingly be smaller by
  up to $-0.4$ dex and the data points move to smaller $b/r_s$ values,
  but because this is not a systematic shift we cannot make a global
  correction to the data in Fig.~\ref{fig:plot2}. 

Some of the scatter seen in Fig.~\ref{fig:plot2} may be caused by DLA
galaxies observed with a spectral resolution that only allows the
emission lines to be marginally resolved and $\sigma_{\mathrm{em}}$ is
therefore uncertain.  Another effect for the data point with the
highest \vsig\ value (from DLA2233+131 at $z=3.151$), is that it has
an unusually small $\sigma_{\mathrm{em}}$. At redshifts $z>3$ the
stellar-mass Tully-Fisher relation has a much larger scatter compared
to at lower redshifts \citep{christensen17}, and the small
$\sigma_{\mathrm{em}}$ may reflect this breakdown of the TF
relation. Therefore the normalisation by $\sigma_{\mathrm{em}}$ may
not be valid at $z>3$.

In addition, the impact parameters are measured in projection, which
could in principle move the data points in Figures~\ref{fig:plot2} and
\ref{fig:plot3} slightly to the right. DLAs likely consist of multiple
individual cloud components that each have different physical
distances from the halo centre, and for a uniform distribution along
the line of sight, the average distance will be dominated by
components close to the plane of the sky, and the correction to the
measured impact parameter will therefore be small.  Occasionally, DLA
systems have a single absorption component, which is significantly
offset from the rest of the components. If it happens to contribute by
more than 5\% of the total optical depth \v90 will be severely
affected. Possibly this single component does belong to the DLA
system, e.g. in a high velocity cloud along the line of sight, or
occasionally to a rotating disk \citep{prochaska97}, or it could be
otherwise unrelated. In this work we use \v90 reported in the
literature along with a few additional values measured from archive
data. For a single one of the DLAs (towards Q0153+0009) we find a
significantly smaller \v90 value than \cite{meiring09}, who determined
a higher value from a weak component offset in velocity compared to
the main bulk of components.

\subsection{Observed and simulated DLA velocity widths}
In order to use the DM models to predict the projected velocity
dispersions, the measured data need to have several trace particles
along the line of sight for a proper comparison.  The observed DLA
systems are likely contained in few gas clouds that are confined
spatially within the galaxy halo, and the global velocity widths of
these gas clouds represent the dynamical motions. In this work we
assume that the velocities represented by \v90 is a measure of the
projected velocity dispersion, and that the components in a DLA system
trace multiple individual clouds along the quasar line-of-sight
\citep[see also the discussion in ][]{moller19}.  Numerical
simulations have investigated the spatial location of these individual
components finding that the high density DLA absorption systems
typically trace a path length of $\sim50-100$ kpc depending on how
large \v90 is \citep{bird15}. As the halo masses in
Table~\ref{tab:halomass} imply virial radii in the range $40-230$ kpc,
similar in size to the simulated objects, we assume that the observed
DLAs probe similar path length through the halos.

The earliest simulations were unable to reproduce the large \v90
\citep{haehnelt98,pontzen08}, but by including feedback from supernova
explosions, modern numerical simulations are better able to reproduce
the observed \v90 distributions \citep{cen12,bird15}. The observed
\vsig\ values have a few values above what can be explained from
simple dark matter plus baryon models hinting that additional velocity
components to \v90 are needed for some of the systems. The data fits
the simulated $\v90/V_{\mathrm{vir}}$ for DLAs remarkably well,
suggesting that the observations do indeed trace gaseous components in
galaxy halos that have been affected by feedback effects. By comparing
the actual measured \vsig\ with the expected value from models alone,
we may derive how much of the DLA gas have been affected by such
feedback mechanisms.

In a future investigation, we will expand this investigation by
analysing simulated DLAs at different redshifts instead of the single
$z=3$ model used here. It would also be interesting to investigate the
dependence of $\v90/V_{\mathrm{vir}}$ with impact parameters in
simulations in order to compare with the observed trends.

\section*{Acknowledgements}
We thank Steen H. Hansen and Radek Wojtak for useful discussions.  LC
and HR are supported by YDUN grant DFF 4090-00079. KEH acknowledges
support by a Project Grant (162948--051) from The Icelandic Research
Fund. The Cosmic Dawn Center is funded by the DNRF.


\bibliographystyle{apj}
\bibliography{ms_v3}
\end{document}